\documentclass[
reprint,
amsmath,amssymb, 
prl, 
jmp,
floatfix,
superscriptaddress,
]{revtex4-2}
\usepackage{xcolor}
\usepackage{graphicx}
\usepackage{dcolumn}
\usepackage{bm}
\usepackage{physics}
\usepackage{url}

\newcommand{\LBL}{Chemical Sciences Division, Lawrence Berkeley National Laboratory, Berkeley, California 94720, USA}
\newcommand{\UCBchem}{Department of Chemistry, University of California, Berkeley, California 94720, USA}
\newcommand{\UCBphys}{Department of Physics, University of California, Berkeley, California 94720, USA}
\newcommand{\Mainz}{Max Planck Institute for Polymer Research, Mainz 55128, Germany}
\newcommand{\Bologna}{Università di Bologna - Alma Mater Studiorum, Via Piero Gobetti 85, 40129 - Bologna, Italy}
\newcommand{\PNNL}{Physical and Computational Sciences Directorate, Pacific Northwest National Laboratory, Richland, WA 99352, USA}
\newcommand{\UCI}{Department of Chemistry and Department of Physics and Astronomy, University of California, Irvine, California 92697-2025, USA}
\newcommand{\UW}{Department of Chemistry, University of Washington, Seattle, WA 98195, USA}

\begin{document}

\title{Tracing long-lived atomic coherences generated via molecular conical intersections}

\author{Patrick Rupprecht}
\email{prupprecht@lbl.gov}
\altaffiliation{These authors contributed equally to this work.}
\affiliation{\UCBchem}
\affiliation{\LBL}
\author{Francesco Montorsi}
\email{francesco.montorsi4@unibo.it}
\altaffiliation{These authors contributed equally to this work.}
\affiliation{\Bologna}
\author{Lei Xu}
\affiliation{\PNNL}
\author{Nicolette G. Puskar}
\affiliation{\UCBchem}
\affiliation{\LBL}
\author{Marco Garavelli}
\affiliation{\Bologna}
\author{Shaul Mukamel}
\affiliation{\UCI}
\author{Niranjan Govind}
\affiliation{\PNNL}
\affiliation{\UW}
\author{Daniel M. Neumark}
\email{dneumark@berkeley.edu}
\affiliation{\UCBchem}
\affiliation{\LBL}
\author{Daniel Keefer}
\email{keeferd@mpip-mainz.mpg.de}
\affiliation{\Mainz}
\author{Stephen R. Leone}
\email{srl@berkeley.edu}
\affiliation{\UCBchem}
\affiliation{\LBL}
\affiliation{\UCBphys}

\date{\today}

\begin{abstract}
Accessing coherences is key to fully understand and control ultrafast dynamics of complex quantum systems like molecules. 
Most photochemical processes are mediated by conical intersections (CIs), which generate coherences between electronic states in molecules. 
We show with accurate calculations performed on gas-phase methyl iodide that CI-induced electronic coherences of spin-orbit-split states persist in atomic iodine after dissociation. 
Our simulation predicts a maximum magnitude of vibronic  coherence in the molecular regime of 0.75\% of the initially photoexcited state population. 
Upon dissociation, one third of this coherence magnitude is transferred to a long-lived atomic coherence where vibrational decoherence can no longer occur.
To trace these dynamics, we propose a table-top experimental approach---heterodyned attosecond four-wave-mixing spectroscopy (Hd-FWM).
This technique can temporally resolve small electronic coherence magnitudes and reconstruct the full complex coherence function via phase cycling. 
Hence, Hd-FWM leads the way to a complete understanding and optimal  control of spin-orbit-coupled electronic states in photochemistry.
\end{abstract}

\maketitle

In the evolution of quantum systems, the population of specific quantum states is relevant as well as the distinct phase relation between them.
This phase relation or quantum coherence has broad applications in physics \cite{streltsov2017colloquium} and chemistry \cite{Schultz2024} and is key for cutting-edge applications like quantum computing \cite{xi2015quantum}.
Coherence between excited electronic states in molecules can be initiated externally through interaction with an ultrafast laser pulse \cite{timmers2019disentangling, saito2019real, kobayashi2022characterizing} or internally via coupled electron-nuclear dynamics on the femtosecond timescale \cite{Kowalewski2017}. 
The latter case is facilitated by wave-packet (WP) splitting at conical intersections (CIs). 
These are singular regions of the potential energy surface (PES) where at least two electronic states are energetically degenerate.
CIs are crucial for most photochemical reactions \cite{domcke2004conical,schultz2004efficient,polli2010conical,tiwari2013electronic,schreier2015early,musser2015evidence,schuurman2018dynamics,wolf2019photochemical}.  
At CIs the initial WP branches into the two intersecting electronic states. 
As both resulting WPs originate from the same initial state, they can exhibit a significant degree of coherence depending on the molecular symmetry \cite{Neville2022}. 
Moreover, since electronic and nuclear dynamics around CIs are strongly coupled, the resulting coherence in the molecule is refereed to as \emph{vibronic} coherence.
The temporal evolution of the vibronic coherence and its vanishing due to intra-molecular vibrations are important components for a complete understanding and potential control of nonadiabatic molecular dynamics. 
It is challenging, however, to directly link measurement signatures to vibronic coherence via standard experimental schemes such as attosecond x-ray transient-absorption spectroscopy (TAS) \cite{cavaletto2023attosecond}.
In recent years, multiple alternative experimental approaches have been proposed and theoretically discussed \cite{bennett2018monitoring,rouxel2021signatures}, one notably labeled TRUECARS \cite{kowalewski2015catching}.
These approaches require large-scale user facilities, such as x-ray free-electron lasers \cite{mcneil2010x} or ultrafast electron diffraction \cite{weathersby2015mega} setups.\\
In this letter, we theoretically investigate the vibronic coherence evolution in gas-phase CH$_3$I molecules after excitation with an ultraviolet (UV) pulse.
Here, the initial photoexcited WP bifurcates at a CI to propagate on two dissociative PESs, which ultimately results in a ground-state or a spin-orbit-excited iodine fragment.
Surprisingly, one third of the maximum coherence magnitude spawned at the molecular CI is maintained in the atomic limit where no vibrational decoherence can occur.
In order to trace the evolution of electronic coherence from the molecular to the atomic regime, we propose and model a novel table-top experimental technique: heterodyned attosecond four-wave-mixing spectroscopy (Hd-FWM).
Theoretical benchmarking of the anticipated Hd-FWM signal demonstrates its direct sensitivity to weak coherences and its capability to monitor the complete complex coherence function.

\begin{figure}
\includegraphics[width=0.45\textwidth]{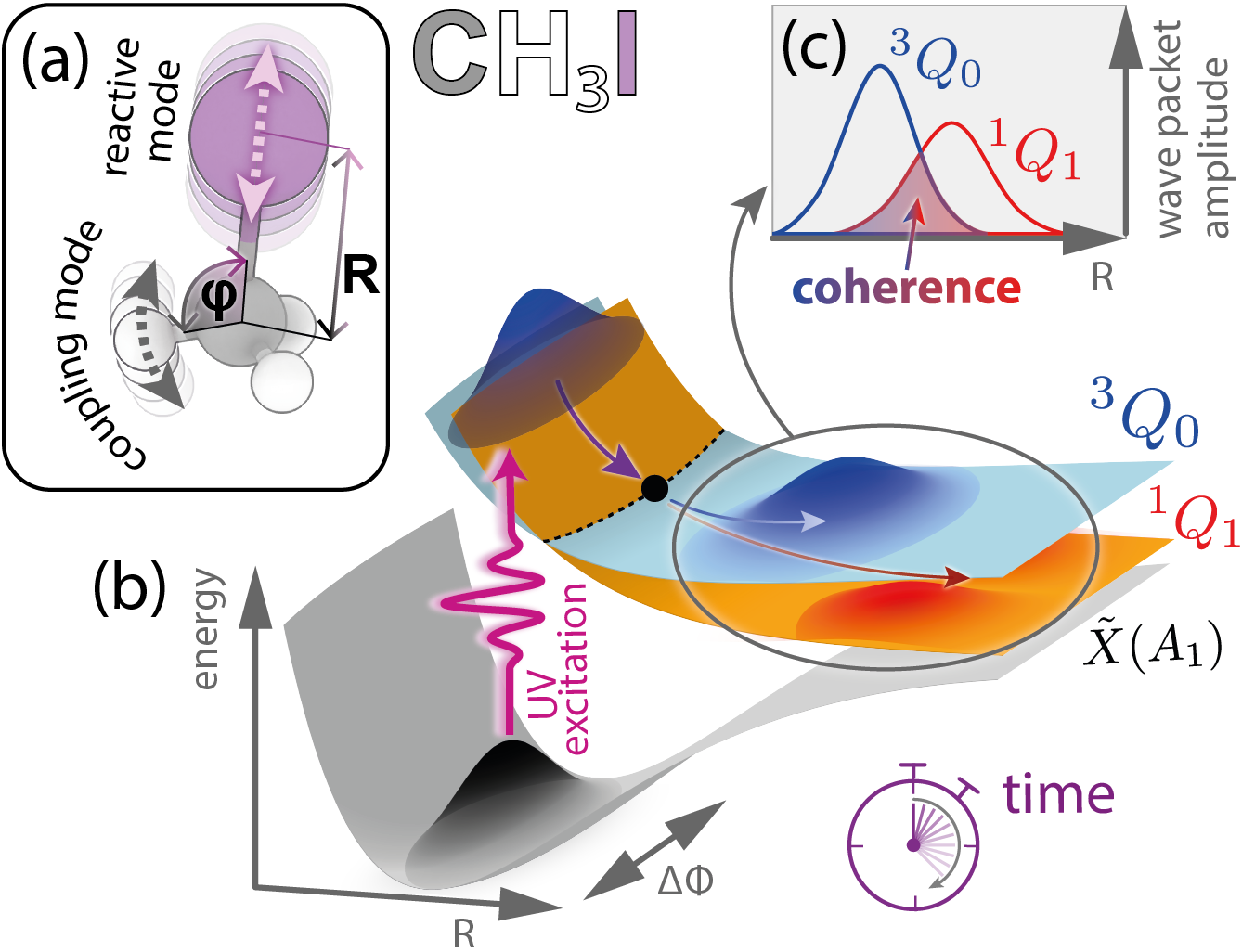}
\caption{\label{fig:wide} Electronic coherence in the photodynamics of CH$_3$I. (a) Molecular structure of CH$_3$I with the indicated reactive vibrational mode $R$ (C--I bond length) and coupling vibrational mode $\varphi$ (H--C--I bending angle) used to model the PESs. (b) Photodynamics along the respective PESs: A WP is excited from the $\widetilde{X}(A_1)$ ground state to the $^3Q_0$ excited-state PES via a UV pulse. After some time, the WP reaches the CI region (dotted line/black dot) where a part of the WP is transferred to the $^1Q_1$ PES. (c) After further propagation, the coherence is determined via the wave-function overlap of both WPs.}
\end{figure}

Photoinduced bond cleavage in CH$_3$I [along $R$ coordinate in Fig.~1(a)] is a paradigm for photodissociation in polyatomic molecules. 
Consequently, this system has been the subject of extensive experimental \cite{imre1984chemical, zhong1998femtosecond, de2008detailed, attar2015direct, drescher2016communication, chang2021mapping, colaizzi2024few} and theoretical \cite{Amatatsu1996,Xie2000,Evenhuis2011,wang2019combined} studies. 
UV excitation around 260\,nm triggers I(5p)$\rightarrow\sigma^*$ electronic transitions in CH$_3$I, which result in the $^3$Q$_0$ and $^1$Q$_1$ spin-orbit-split states \cite{roehl1997temperature,eppink1999energy}. 
Both display steep dissociative PESs as illustrated in Fig.~1(b) that lead to the $j=\{1/2,3/2\}$ states of the atomic iodine fragment. 
Laser excitation initiates the dynamics predominantly on the $^3$Q$_0$ PES \cite{chang2021mapping,Evenhuis2011}, which is accessed via a one-UV-photon transition from the $\Tilde{X}(A_1)$ ground state. 
During dissociation, a $^3$Q$_0$/$^1$Q$_1$ CI is encountered after about 15\,fs \cite{chang2021mapping}. 
This results in the generation of an vibronic coherence that is determined by the $^3$Q$_0$/$^1$Q$_1$ WP overlap $\braket{\chi_{^3Q_0}(t)}{\chi_{^1Q_1}(t)}$ [see Fig.~1(c)].\par
We model this phenomenon using exact nuclear quantum dynamics (QD) in reduced dimensions as implemented in the QDng software \cite{https://doi.org/10.5281/zenodo.10944496}.
In particular, we employ an effective two-dimensional  Hamiltonian comprising the diabatic $^3$Q$_0$ and $^1$Q$_1$ states, which reproduces the results of the full nine-dimensional \textit{ab-initio} model previously reported by Amatatsu \textit{et al.} \cite{Amatatsu1996} and further corrected by Xie \textit{et al.} \cite{Xie2000}. 
The two nuclear degrees of freedom spanning the system Hamiltonian were chosen to encompass a large cross section of the $^3$Q$_0$/$^1$Q$_1$ branching space. 
As illustrated in Fig.~1(a) these coordinates are the symmetry preserving (\emph{reactive}) C$-$I stretching ($R$) and the symmetry breaking (\emph{coupling}) H$-$C$-$I bending ($\varphi$). 
This QD setup provides population dynamics and $j=\{1/2,3/2\}$ branching ratios (Fig.~S2 in the Supplemental Material) in good agreement with previous theoretical studies \cite{Evenhuis2011,wang2019combined}. \par
The vibronic coherence established at the CI is probed by spectroscopically targeting the 4d $\rightarrow$ 5p core-to-valence transition of iodine \cite{o1982absorption}. 
This results in a pair of spin-orbit-split core-excited states, $\sigma^*(4d_{5/2})^{-1}$ and $\sigma^*(4d_{3/2})^{-1}$. 
These states are computed by many-body all-electron quantum chemistry (QC) calculations using the multi-reference configuration interaction (MRCI) \cite{Knowles_1988,Knowles_1992} method including spin-orbit coupling effects as implemented in the MOLPRO \cite{Molpro12} package.
More details about the QC simulations are given in SM Sec.~III.

We propose the Hd-FWM technique for directly measuring the electronic coherence evolution in CH$_3$I in a time-resolved manner.
Noncollinear attosecond FWM spectroscopy is a well established table-top experimental approach that targets the $\chi^{(3)}$ nonlinear response of highly-excited states in atoms \cite{cao2016noncollinear, fidler2019nonlinear, marroux2018multidimensional, fidler2019autoionization, puskar2023measuring, gaynor2023nonresonant, rupprecht2024extracting}, molecules \cite{cao2018excited, lin2021coupled, fidler2022state}, and solid-state systems \cite{gaynor2021solid} by mixing a broadband XUV pulse with two few-cycle near infrared (NIR) or visible (VIS) pulses. 
The attosecond XUV pulse is produced via high-order harmonic generation \cite{corkum1993plasma,schafer1993above,krausz2009attosecond} and shares the same driving laser as the NIR/VIS pulses, hence enabling precise timing and control over the pulse characteristics.
So far, attosecond FWM spectroscopy has been used to measure lifetimes of highly-excited states in a quantum-state specific manner \cite{puskar2023measuring,rupprecht2024extracting} as well as to characterize electronic coherences in atomic systems \cite{cao2016near,gaynor2023nonresonant}.\par
The novelty of the proposed approach consists of interfering two FWM signals, which enhances the sensitivity such that this technique can unambiguously measure small vibronic coherence magnitudes in molecules.
Fig.~2 shows the general experimental scheme: The ultrashort UV pump pulse centered at 260\,nm photoexcites CH$_3$I. 
\begin{figure}
\includegraphics[width=0.5\textwidth]{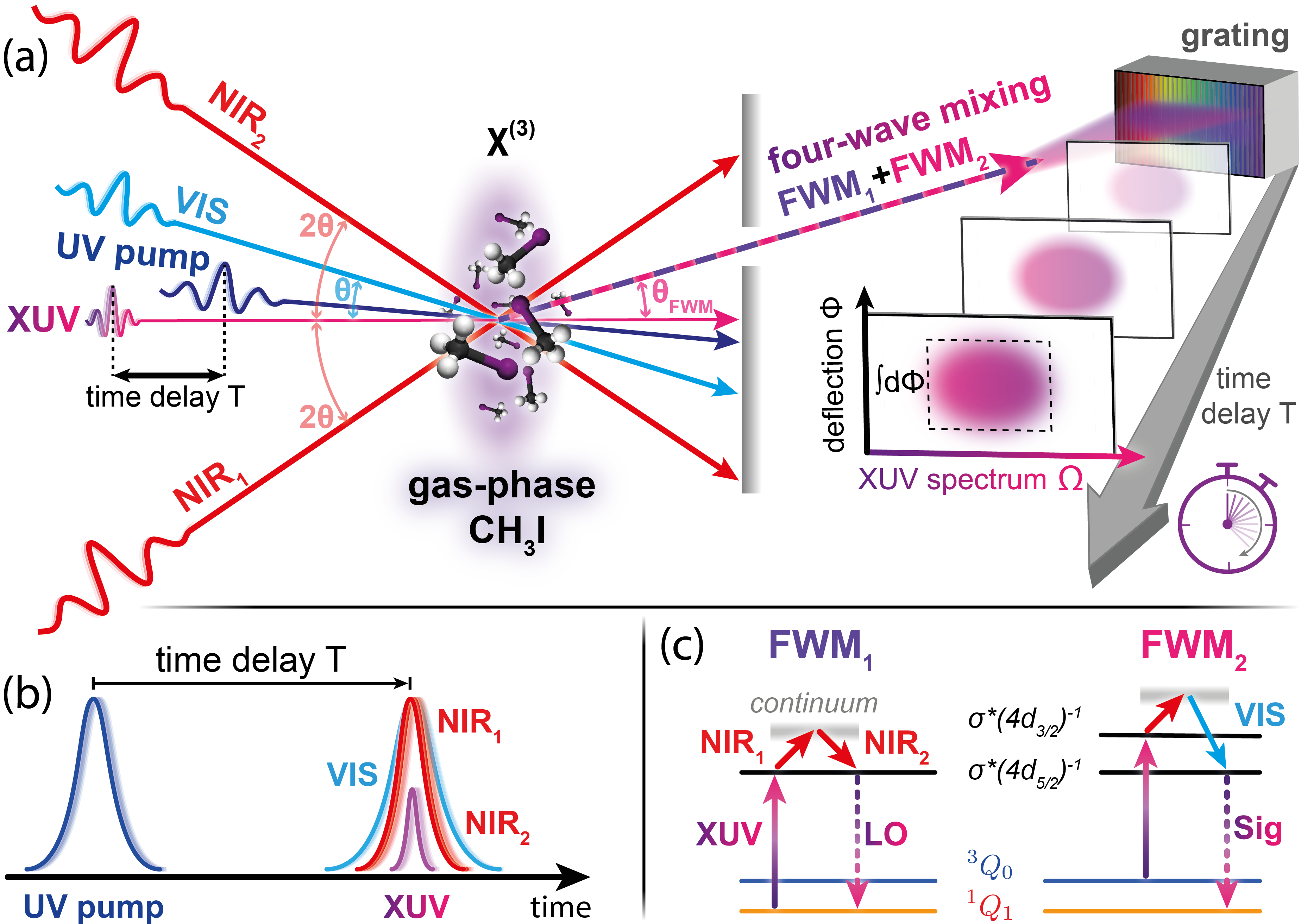}
\caption{Heterodyned attosecond four-wave-mixing spectroscopic scheme. (a) After photoexcitation via the UV pump pulse, four time-delayed pulses interact with gas-phase CH$_3$I. First, the broadband XUV pulse, the NIR$_1$, and the NIR$_2$ pulses result in the FWM$_1$ signal via a $\chi^{(3)}$ process. 
Additionally, the XUV, NIR$_1$, and VIS pulses result in the FWM$_2$ signal. By choosing the correct noncollinear geometry, FWM$_1$ and FWM$_2$ overlap spectrally and spatially on a CCD chip of the XUV spectrograph. (b) The measurement variable is the time delay between the UV and all other pulses. (c) Hd-FWM quantum pathways: NIR$_1$ and NIR$_2$ couple to the continuum in a Raman process and result in a degenerate FWM$_1$ emission, which acts as local oscillator (LO). FWM$_2$ utilizes a Raman process via NIR$_1$ and the VIS pulse to result in the coherence-encoding signal (Sig).}
\end{figure}
Four pulses then act at a variable time delay and contribute to the FWM signals:
A broadband XUV pulse interacts with NIR$_1$ and NIR$_2$ ($\lambda_{NIR} = 800$\,nm) in the gas-phase CH$_3$I sample resulting in the FWM$_1$ signal. 
Due to the noncollinear beam geometry (XUV-NIR angle $\pm 2\Theta$) and the $\Lambda$-coupling scheme [see Fig.~2(c)], FWM$_1$ propagates with a noncollinear angle $\Theta_{FWM}$ with respect to the XUV input beam in accordance with the phase-matching condition. 
This resulting noncollinear emission angle $\Theta_{FWM}$ enables the spatial separation from the transmitted input pulses.
The second signal FWM$_2$ results from the interaction of XUV, NIR$_1$ and a pulse at the second harmonic of 800\,nm (VIS; $\lambda_{VIS} = 400$\,nm).
By choosing the XUV-VIS angle to be $\Theta$, FWM$_1$ and FWM$_2$ share the same noncollinear angle and hence are spatially overlapped at the spectrograph. 
Spectral overlap with FWM$_1$ is evident by the respective coupling pathway of FWM$_2$ in Fig.~2(c):
The XUV-emission transition is in both cases $\sigma^*(4d_{5/2})^{-1} \rightarrow ^1$Q$_1$.
While FWM$_1$ projects the population of the $^1$Q$_1$ PES onto itself via coupling to the $\sigma^*(4d_{5/2})^{-1}$ core-excited PES and acts as a reference field (local oscillator; LO), FWM$_2$ couples the $^3$Q$_0$ and $^1$Q$_1$ PESs and hence encodes the desired coherence signal (Sig). \par
The total intensity $\text{I}_\text{T}$ detected by the XUV spectrograph at $\Theta_{FWM}$ as a function of the time delay $T$ results from the superposition of the LO ($\text{E}_{\text{LO}}$) and signal ($\text{E}_{\text{S}}$) electric fields \cite{MukamelBook}:
\begin{equation}
\begin{aligned}
\text{I}_\text{T}(T,\Omega) \propto &  \, \text{I}_\text{LO}(T,\Omega) + \text{I}_\text{S}(T,\Omega) +\\
 & + 2\Re \left[ \text{E}^*_{\text{LO}}(T,\Omega)\text{E}_{\text{S}}(T,\Omega)\right] \ \ .
\end{aligned}
\label{eq:tot_sig}
\end{equation}
\noindent Here, $\text{I}_\text{LO}$ and $\text{I}_\text{S}$ are the intensities of the LO and signal components respectively (i.e., $|\text{E}_\text{LO}|^2$ and $|\text{E}_\text{S}|^2$) while $\Omega$ is the FWM emission photon energy. 
The LO is directly sensitive to the $^1Q_1$ population while the signal detects the much weaker electronic coherence.
This results in $\text{E}_\text{LO}>>\text{E}_\text{S}$. 
Therefore, the $\text{I}_\text{S}$ term in Eq.~\eqref{eq:tot_sig} can be neglected. 
Moreover, $\text{I}_\text{LO}$ is easily measured by blocking the VIS pulse. 
Consequently, by subtracting $\text{I}_\text{LO}$ from $\text{I}_\text{T}$, the FWM interference is isolated yielding the Hd-FWM measured intensity signal 
\begin{equation}
\begin{aligned}
\text{I}_\text{Hd}(T,\Omega) \propto 2\Re \left[ \text{E}^*_{\text{LO}}(T,\Omega)\text{E}_{\text{S}}(T,\Omega)\right] \ .
\end{aligned}
\label{eq:HD}
\end{equation}

\noindent Both $\text{E}_\text{LO}$ and $\text{E}_\text{S}$ are evaluated here using third-order perturbation theory. That is achieved by expanding the FWM$_1$ and FWM$_2$ response functions as a series of Feynman diagrams \cite{MukamelBook} with each one corresponding to a specific quantum pathway connecting valence and core excited states (see Fig.~S5 in the SM).
This rationalizes the Hd-FWM process as the interference between two dominating coupling pathways as depicted in Fig.~2(c). 
Other possible coupling pathways are not phase-matched at $\Theta_{FWM}$, emit at a different XUV photon energy or are negligible due to transition-dipole-moment considerations (see SM Sec.~IVb).\par

\begin{figure}
\includegraphics[width=0.45\textwidth]{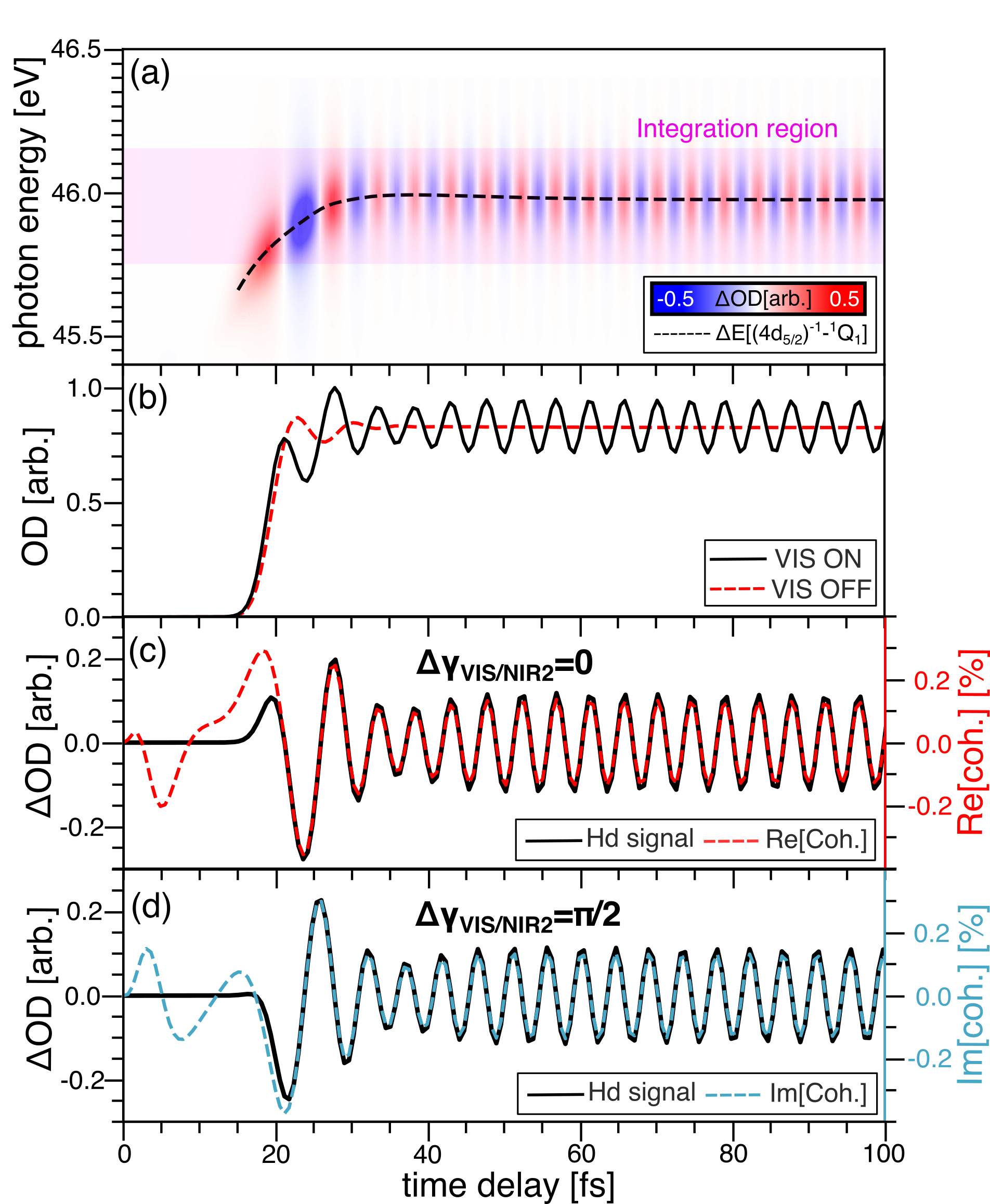}
\caption{Simulated Hd-FWM electronic coherence signal. (a) Hd-FWM spectrogram as function of the FWM emission energy $\Omega$ and \emph{pump}-\emph{probe} time delay $T$. The black dashed line depicts the time evolution of the $\sigma^*(4d_{5/2})^{-1}$-$^1Q_1$ energy gap during the dynamics. (b) Simulated FWM trace with ($\text{I}_\text{T}(T,\Omega)$; black solid line) and without ($\text{I}_\text{LO}(T,\Omega)$; red dashed line) the VIS pulse highlighting the formation of FWM interference. (c) Oscillating component $\text{I}_\text{Hd}(T,\Omega)$ of (b) matching the real part of the electronic coherence (red dashed line). (d) same as (c) but applying a relative CEP offset of $\pi/2$ between the VIS and the NIR$_2$ pulses. Now, the Hd signal traces the imaginary part of the coherence (dashed light-blue line) granting access to the full complex coherence function. Traces (b), (c), and (d) are obtained by integrating the signal in the photon-energy window displayed in (a).}
\end{figure}

The  Hd-FWM simulation results in the $\text{I}_\text{T}(T,\Omega)$ spectrogram that is displayed in Fig.~3(a). 
We observe an oscillating feature in the spectral region between 45.5 and 46.5\,eV, which rises at around 15\,fs, at the conical intersection. 
Here, the $\chi_{^3Q_0}$ WP splits leading to the simultaneous occurrence of both FWM$_1$ and FWM$_2$, which is required to observe the Hd-FWM signal.
Moreover, the centroid of the signal displays a transient blue shift tracking the increasing $\sigma^*(4d_{5/2})^{-1}$-$^1Q_1$ energy gap during the dynamics [black dashed line of Fig.~3(a)]. 
Such a behavior can be understood by benchmarking valence against core-excited-state PESs (see SM Fig.~S4).   
If the relative carrier-envelope phase (CEP) between the NIR$_2$ and VIS pulses is zero, the intensity oscillation recorded in the Hd-FWM signal matches the real part [see Fig.~3(b)] of the coherence spawned at the molecular CI, therefore providing a direct measure of the coherence evolution [see Fig.~3(c)].
Alternatively, if the relative CEP is set to $\pi/2$, the Hd-FWM maps the imaginary part of the coherence [see Fig.~3(d)]. 
This demonstrates that Hd-FWM uniquely provides access to the complete complex coherence by probing separately its real and imaginary part via a phase-cycling approach known from other multidimensional spectroscopy schemes \cite{tian2003femtosecond}. 
As XUV and NIR$_1$ pulses are simultaneously involved in both FWM$_1$ and FWM$_2$, their CEPs are not cycled since they do not affect the relative phase between the $\text{E}_\text{LO}$ and the $\text{E}_\text{S}$ field.\par
Notably, the anticipated vibronic coherence magnitude in CH$_3$I is minimal, i.e. about 0.75\% of the photoexcited initial $^3Q_0$ state population. 
This is a consequence of the different symmetries displayed by $^1Q_1$ and $^3Q_0$, which respectively belong to the E(A$^\prime$) and A$_1$ irreducible representations of the C$_{3v}$ point group \cite{Evenhuis2011}. 
As previously found by Neville \textit{et al.} \cite{Neville2022} in a similar case, this differing symmetry affects the relative parity of the two nuclear WPs with respect to the coupling mode $\varphi$ thus making them nearly orthogonal. 
Specifically, the $\chi_{^3Q_0}$ WP is an even function showing a maximum around the equilibrium H$-$C$-$I bending angle, whereas the $\chi_{^1Q_0}$ WP, which is an odd function, has a minimum (see Fig.~1(b) and the WP evolutions provided in the WP movie reported as electronic SM).
The strict symmetries of the WPs, however, are broken by anharmonic contributions to the $^1Q_1$ and $^3Q_0$ PESs along the bending angle coordinate $\varphi$ (see Eqs.~(S2) and (S3) in the SM). 
This prevents the WP overlap $\braket{\chi_{^3Q_0}(t)}{\chi_{^1Q_1}(t)}$ from vanishing completely.
Interestingly, by leveraging on the interference with the intense FWM$_1$ signal, even such a weak coherence is brought towards reasonable detection limits. 
As a demonstration of the sensitivity of the technique, Fig.~3(b) benchmarks the total FWM intensity [i.e., $\text{I}_{\text{T}}$ in Eq.~\eqref{eq:tot_sig}] against the LO background obtained by blocking the VIS pulse [$\text{I}_{\text{LO}}$ in Eq.~\eqref{eq:tot_sig}]. 
In particular, the FWM interference-fringe contrast is around 20\% of the LO intensity. 
This represents a clear advantage of the heterodyne approach over homodyne detected FWM and conventional TAS, neither of which benefit from such an improved sensitivity and unambiguous signal relation to electronic coherences. \par
One major insight from the simulated coherence presented in Fig.~3 is that, surprisingly, the coherence spawned at the molecular CI does not dephase completely when the iodine atom is fully dissociated from the rest of the molecule. 
Instead, Figs.~3(c)-(d) show that for T$>80$\,fs the electronic coherence exhibits a constant oscillation with a 4.4\,fs periodicity. 
This value corresponds to the spin-orbit splitting between the $j=\{1/2,3/2\}$ states of iodine of $E_{SO} = 0.94$\,eV \cite{o1996trends}, suggesting that the original $^1Q_1$-$^3Q_0$ electronic coherence, generated in the molecule, is observed at the atomic I fragment.\par
To further elucidate the evolution of the molecular complex coherence to the atomic regime we visualize its magnitude and phase in the Euler representation (see Fig.~4). 
Here, the initial coherence magnitude converges to constant values (about $1/3$ of its maximum) for T$>80$\,fs, while the coherence phase converges to a constant slope. 
This is expected in the atomic limit where nuclear motion does not influence the electronic coherence and no vibrational dephasing can occur. 
The CH$_3$I photodissociation time of about 80\,fs, extrapolated with this analysis, agrees with reported experimental values \cite{corrales2014structural}.
\begin{figure}
\includegraphics[width=0.45\textwidth]{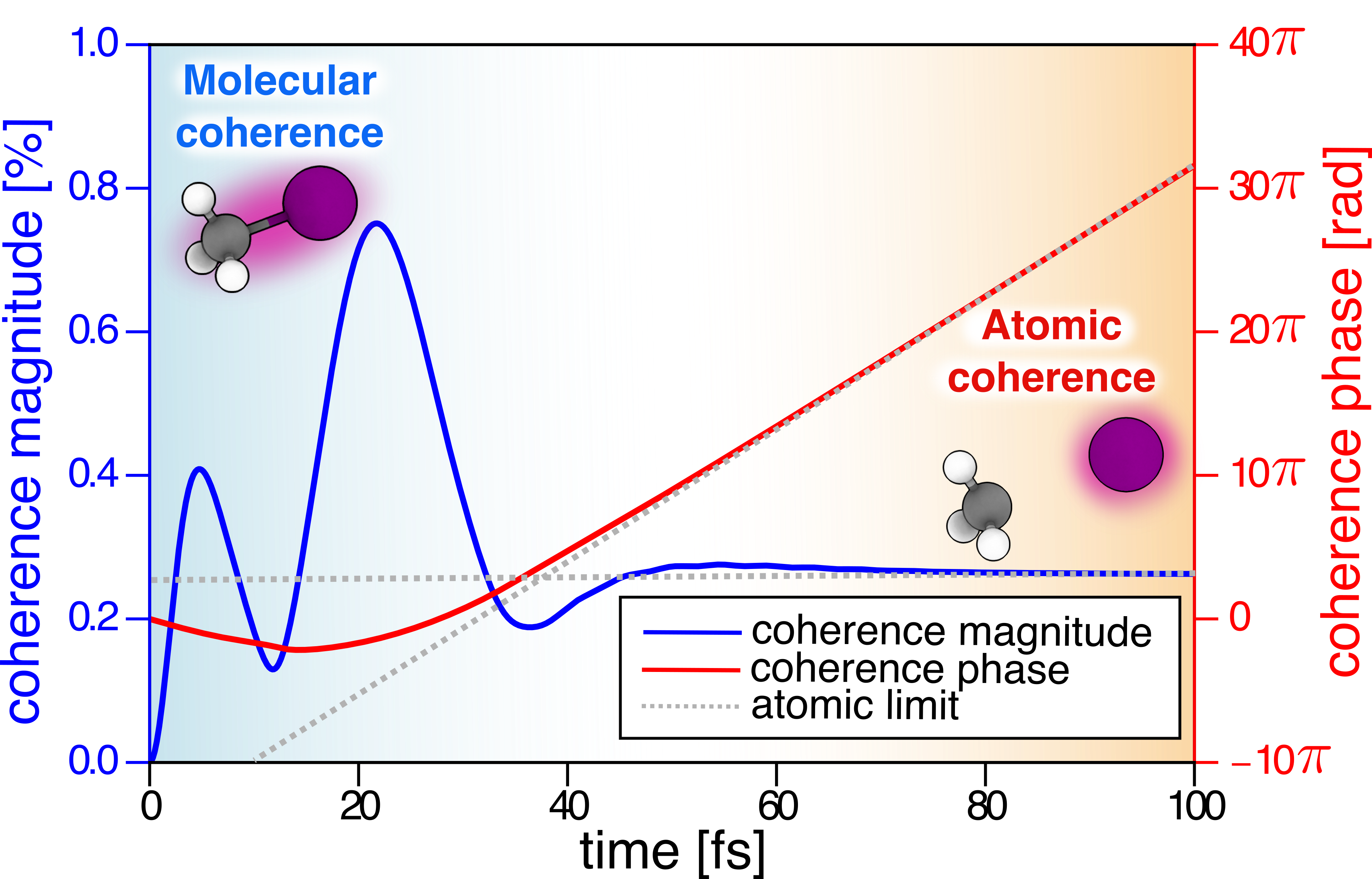}
\caption{Evolution of molecular coherence to the atomic regime. For early times after UV excitation ($< 40$\,fs), the vibronic coherence magnitude (blue) shows large time-dependent modulations, while it converges to a constant value for later times ($> 80$\,fs). The coherence phase (red) converges to a linear evolution for these later times. This convergence of the coherence magnitude and phase is a consequence of transitioning from a molecular to a dissociated atomic regime.}
\end{figure}
The transition from the molecular to the atomic regime starts at 40\,fs when the WPs travel away from the CI region. 
Due to the steeper slope of the $^1Q_1$ compared to the $^3Q_0$ PES, the centroid of the $\chi_{^1Q_1}$ WP moves faster along R than the $\chi_{^3Q_0}$ WP. These different velocities are compensated by the broadening of the two WPs, surprisingly leading to a constant WP overlap and preventing a complete dephasing of the molecular coherence upon dissociation. \par
The fast oscillations in the atomic limit impose high temporal resolution requirements on the experiment.
First, the actinic photoexcitation will be achieved with sub-5\,fs UV pulses \cite{graf2008intense, galli2019generation, colaizzi2024few, travers2019high, reduzzi2023direct} clearly separating the photoexcitation regime from the WP propagation through the CI.
Second, the Hd-FWM signal occurs after passage through the CI (T~$>15$\,fs; see Fig.~3) and hence the instrument response function (IRF) out of temporal overlap with the actinic UV pulse is decisive.
While the NIR (e.g., $\tau_{NIR} < 5$\,fs \cite{cavalieri2007intense, timmers2017generating,nagy2021high}) and VIS pulses (e.g., $\tau_{VIS} < 25$\,fs \cite{liu2010generation, hu2023clean, ou2024attosecond}) are of femtosecond duration, the isolated attosecond XUV pulse \cite{hentschel2001attosecond,krausz2009attosecond,chini2014generation} acts as a temporal gate via core-level excitation.
These laser pulse sources are readily available in many labs worldwide and lead to a sub-femtosecond Hd-FWM IRF rise time (more details are given in the SM Sec.~V).
They further typically provide a pulse-intensity stability of $< 0.5\%$ for the NIR \cite{TiSa_stab,rupprecht2022ultrafast} and the VIS pulses \cite{hu2023clean} as well as a CEP stability of $< 200$\,mrad \cite{musheghyan20190}. 
The Hd-FWM signal, however, is barely affected by such noise (see SM Sec.~VI).
Due to the heterodyned FWM scheme, only intensity and relative CEP fluctuations of the NIR$_2$ and the VIS pulses contribute to any potential reduction in the contrast of the Hd-FWM interference signal, while noise in the XUV and the NIR$_1$ pulses cancel out. \par
In summary, we show that one third of the maximum vibronic coherence generated in photoexcited CH$_3$I by a CI remains as electronic coherence in the atomic iodine fragments after dissociation.
This allows one to measure coherences spawned at CIs even when the pump-probe measurement delay is long after the WP propagation through the CI.
To measure the evolution of the respective coherence between $^3$Q$_0$ and $^1$Q$_1$ electronic states we present the heterodyned attosecond FWM spectroscopic technique.
Here, two FWM signals that originate from mixing an attosecond XUV pulse with two NIR pulses or with one NIR and one VIS pulse, respectively, can be interfered with each other.
The Hd-FWM signal directly maps the temporal evolution of the vibronic coherence with high sensitivity and sub-femtosecond resolution.
Moreover, it allows one to trace the full complex coherence function by cycling the relative CEP between one NIR and the VIS pulse and therefore enabling a quantum-state tomography.
The Hd-FWM technique as discussed in this letter is based on realistic table-top experimental parameters and is hence relevant and accessible for the ultrafast scientific community.\par
This work frontiers the idea of using photochemistry to prepare long-lived electronic coherence in atoms after molecular dissociation.
As this approach avoids decoherence effects in molecules, it results in robust coherent states of gas-phase atoms and hence constitutes a potential building block for ultrafast quantum information technology \cite{de2007femtosecond, underwood2003switched, koll2022experimental, shobeiry2024emission, bouchard2024programmable}.
Moreover, this work adds to the toolbox of coherent-control of spin-orbit effects \cite{rupprecht2022laser}. \\
\\
\begin{acknowledgments}
We thank Jérémy R. Rouxel for valuable discussions.
This work was performed by personnel and equipment supported by the Office of Science, Office of Basic Energy Sciences through the Atomic, Molecular, and Optical Sciences Program of the Division of Chemical Sciences, Geosciences, and Biosciences of the U.S. Department of Energy (DOE) at Lawrence Berkeley National Laboratory under Contract No. DE-AC02-05CH11231 (P.~R., N.~G.~P., D.~M.~N., and S.~R.~L.) and Contract No. DE-SC0022225 (S.~M. at the University of California, Irvine; F.~M., M.~G. at the University of Bologna, Italy; L.~X. and N.~G. under FWP 72684 at the Pacific Northwest National Laboratory (PNNL)). P.~R. acknowledges funding by the Alexander von Humboldt Foundation (Feodor Lynen Fellowship). N.~G.~P. acknowledges funding from Soroptimist International of the Americas (Founder Region Fellowship). This work was also supported by ERC grant QuantXS, 101116417. The research benefited from computational resources provided by PNNL’s Institutional Computing (PIC) Program, Environmental Molecular Sciences Laboratory (EMSL), a DOE Office of Science User Facility sponsored by the Office of Biological and Environmental Research and located at PNNL, and the National Energy Research Scientific Computing Center (NERSC), a U.S. Department of Energy Office of Science User Facility operated under Contract No. DE-AC02-05CH11231. PNNL is operated by Battelle Memorial Institute for the United States Department of Energy under DOE Contract No. DE-AC05-76RL1830. \par
P.~R. and F.~M. contributed equally to this work. P.~R. and S.~R.~L. conceptualized the project. P.~R., N.~G.~P., D.~M.~N., and S.~R.~L. developed the Hd-FWM scheme. F.~M. and D.~K. conceptualized the response theory. L.~X. and N.~G. conducted the quantum chemistry calculation. F.~M., and D.~K. conducted the Hd-FWM simulation with input from S.~M. and M.~G. . 
N.~G., M.~G., D.~M.~N., S.~R.~L., and D.~K. supervised the work. P.~R. and F.~M. wrote the manuscript with input from all authors.
\end{acknowledgments}

\bibliography{bib}

\end{document}